\documentclass[a4paper,11pt]{article}
\pdfoutput=1 

\usepackage{jinstpub} 
\usepackage{bibentry}
\usepackage{booktabs}

\title{\boldmath Energy Reconstruction of Hadrons in highly granular combined ECAL and HCAL systems}

 \author{Y. Israeli}
 \affiliation{Max-Planck-Institute for Physics,\\Munich, Germany}

\emailAdd{israeli@mpp.mpg.de}

\abstract{
This paper  discusses the hadronic energy reconstruction of two combined electromagnetic and hadronic calorimeter systems using physics prototypes of the CALICE collaboration: the  silicon-tungsten electromagnetic calorimeter (Si-W ECAL) and the  scintillator-SiPM  based analog hadron calorimeter (AHCAL); and the scintillator-tungsten electromagnetic calorimeter (ScECAL) and the AHCAL. These systems were operated in hadron beams at CERN and FNAL, permitting the study of the performance in combined ECAL and HCAL systems.

Two techniques for the energy reconstruction are used, a standard reconstruction based on calibrated sub-detector energy sums, and one based on a software compensation algorithm making use of the local energy density information provided by the high granularity of the detectors. The software compensation-based algorithm improves the hadronic energy resolution by up to 30\% compared to the standard reconstruction. The combined system data show  comparable
 energy resolutions to the one achieved for data with showers starting only in the AHCAL and therefore demonstrate the success of the inter-calibration of the different sub-systems, despite of their different geometries and different readout technologies.
}


\collaboration[c]{on behalf of CALICE collaboration}

\proceeding{CHEF2017\\
  2-6 October 2017\\
  Lyon, France}

\begin{document}
\maketitle
\flushbottom

\section{Introduction}
\label{sec:intro}
 The CALICE collaboration develops highly granular calorimeters for present and future collider experiments. Among the physics prototypes already tested extensively in particle beams are silicon-tungsten (Si-W ECAL) \cite{ECAL}  and scintillator-tungsten (ScECAL) \cite{ScECAL} electromagnetic calorimeters and a scintillator-SiPM  based analog hadron calorimeter (AHCAL) \cite{AHCAL}. 
 In order to evaluate the performance of the calorimeter prototypes in realistic detector configurations,  each electromagnetic calorimeter was installed together with the AHCAL and a scintillator-steel tail catcher and muon tracker (TCMT) \cite{TCMT} and the combined system was  tested in test beam experiments.

 The present paper presents the performance of these combined systems in detecting negative hadron beams with a momentum range of 4\,GeV to 80\,GeV. The data were recorded at CERN in 2007 and FNAL in 2008 with the  Si-W ECAL, AHCAL and TCMT setup (\textit{Si/Scint setup}) and at FNAL in 2009 with the ScECAL, AHCAL and TCMT setup  (\textit{all-Scint setup}).
  
 The data were reconstructed with the so called \textit{standard reconstruction}, where a simple sum of  the contributions of the sub-detectors is computed and with the \textit{software reconstruction} (SC) method, which uses the information of the local energy of each detector cell. More details can be found in CALICE analysis notes: \cite{Oskar} and \cite{Yasmine}.

\section{Energy Reconstruction of Hadrons}
In a combined calorimeter system with different geometries and different readout technologies, the reconstruction of the total energy deposit by a particle requires several calibration and reconstruction steps. 
As a first step, the response of the cells of each sub-detector is calibrated to the units of  MIPs (minimum-ionising particles) with muon beams, as described in detail in \cite{AHCAL}. Only cells with a signal above 0.5 MIP were considered for further
analysis and are called hits.
Following this, calibration factors from MIPs to GeV are determined  for each technologically or geometrically distinct region of the system to equalized the sub-detector signals. In this analysis, the same calibration is applied to the AHCAL and the TCMT due to the similar technologies and identical sampling structure used in these calorimeters \cite{AHCAL, TCMT}. 
For each beam energy, two  calibration factors, one for the ECAL and one for the AHCAL and the TCMT, are determined  with a  $\chi^2$ minimization method using the known test beam energy, $E_\text{beam}$. These calibration factors show a slight energy dependence, with changes on the level of  1\%-3\%  over the energy ranges considered here. A set of global calibration factors is obtained by averaging the factors over all energies for a given beam period.
 After the calibration, the sub-detector energies are summed to determine the particle energy  with the standard reconstruction method. For each  $E_\text{beam}$ the distribution of the reconstructed energies is fitted to determine the mean reconstructed energy  $E_\text{reco}$ and the standard deviation $\sigma_\text{reco}$, where the    energy resolution is given by  $\sigma_\text{reco}/E_\text{reco}$.

Since the sub-detectors in this analysis are non-compensating calorimeters, their response for electromagnetic  showers is typically larger than for hadronic showers. In addition, the electromagnetic fraction of the hadronic showers, from the production of neutral pions, fluctuates from event to event and therefore reduces the energy resolution.
A reconstruction technique which was shown previously to improve the energy resolution for the AHCAL \cite{Frank} is the software compensation method. In this method, different weights are assigned to different calorimeter cells based on their local energy density.

\begin{figure}[h]
\includegraphics[width = 0.495\textwidth]{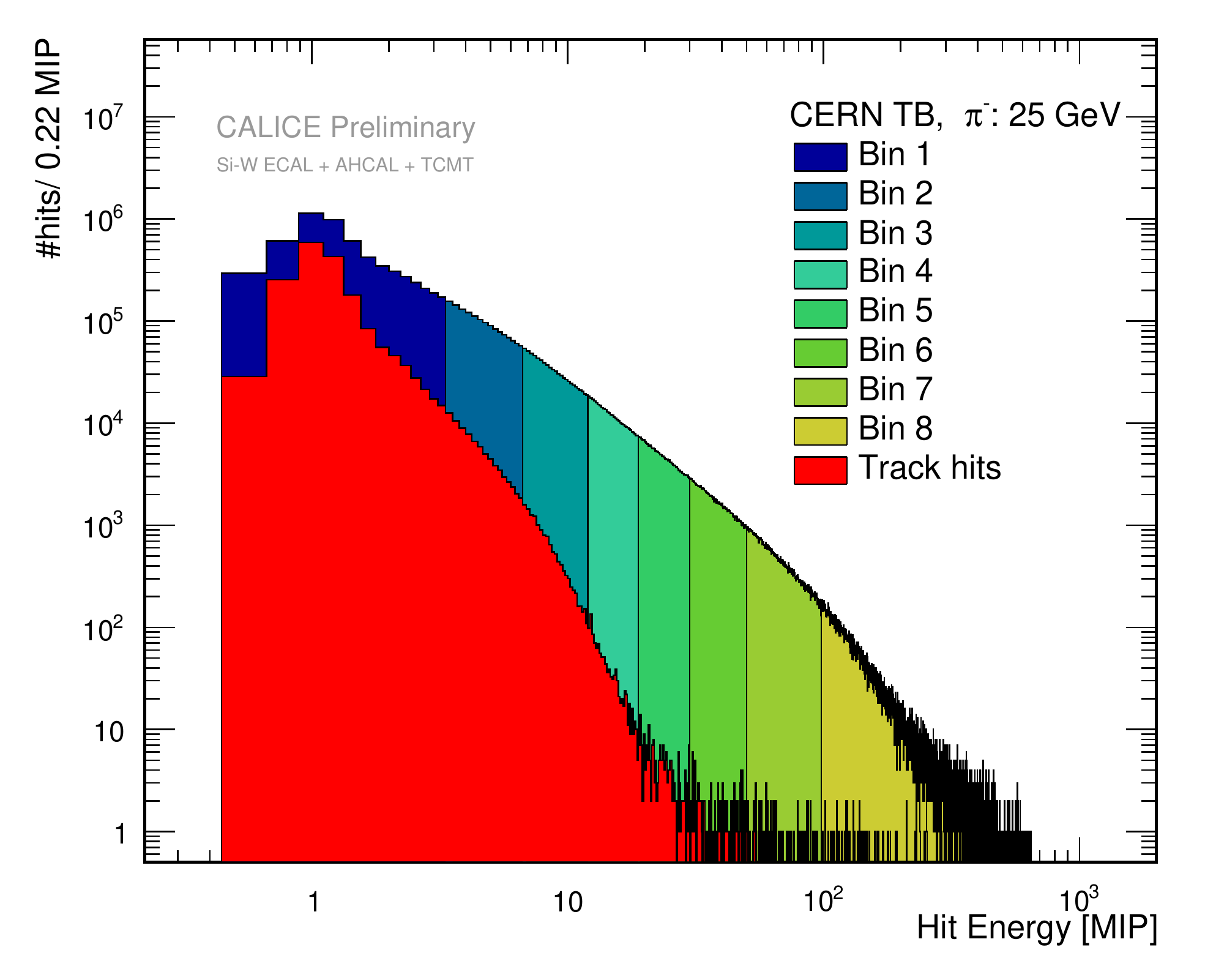} 
\hfill
\includegraphics[width = 0.495\textwidth]{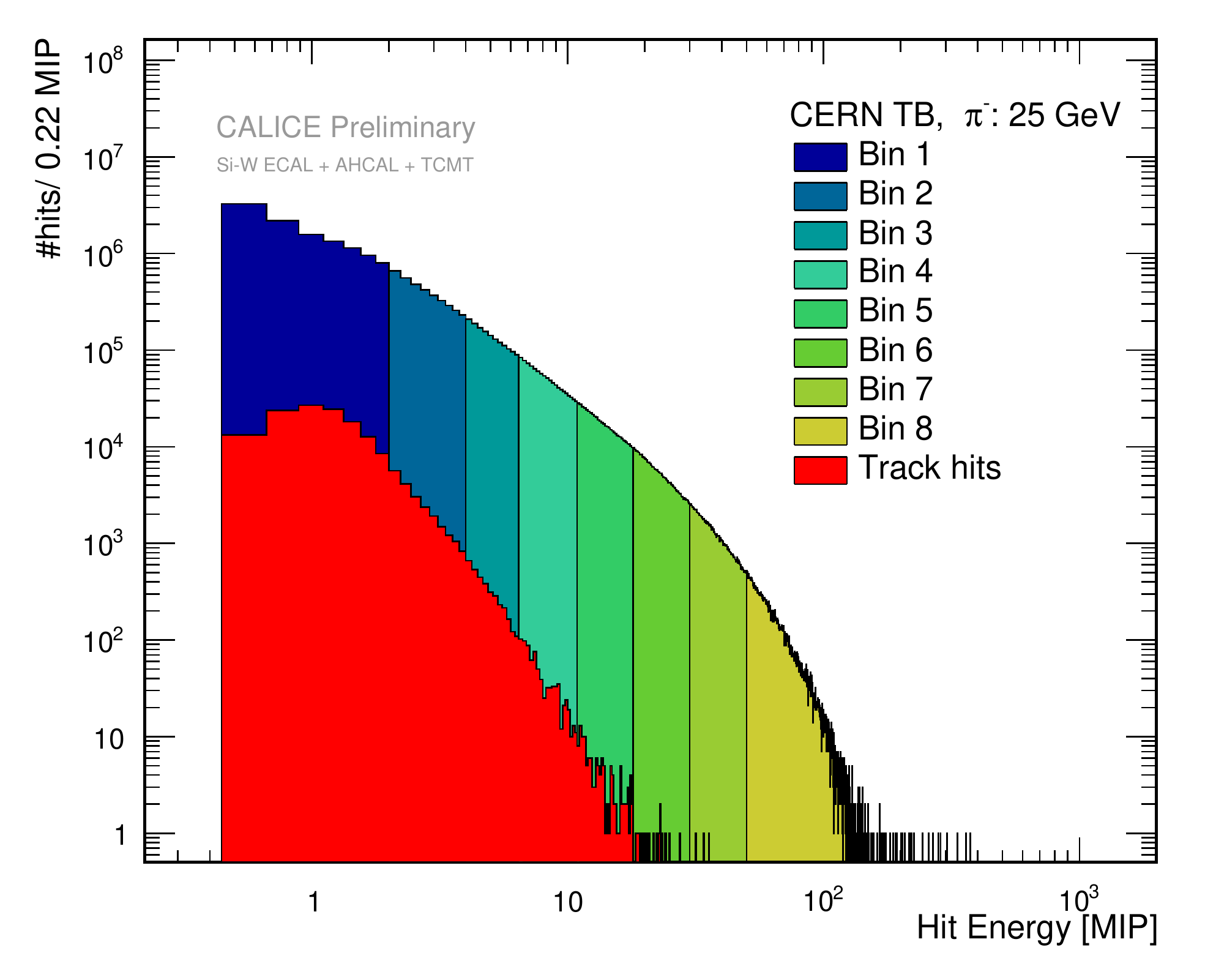} 
\caption{Distribution of the hits of 25 GeV $\pi^-$ of CERN TB in the Si-W ECAL (left) and AHCAL (right). The different colors show the different bins and the track energy depositions.}
\label{fig:bins}
\end{figure}

The SC scheme  in this study includes each of the sub-detectors in the combined system.
The  distribution of hit energies  in each sub-detector is divided into several bins for different energy ranges (excluding the primary track hits): eight bins for the Si-W ECAL and the ScECAL, eight bins for the AHCAL and one bin for the TCMT.  Figure \ref{fig:bins} presents an example of the Si-W ECAL and AHCAL hit distributions  in the Si/Scint setup for 25\,GeV $\pi^-$ in the CERN test beam.  The different colors represent the different bins and the primary track hits (red). 
 
For each bin $i$ ($i$=3-8 for the ECALs and  AHCAL distributions, $i$=1 for TCMT distribution) the total bin energy $E_i$ is calculated, taking into account the appropriated calibration factor for the sub-detector. 
For the first two bins in the  ECALs and the AHCAL,  $E_i$ is given by the number of the hits multiplied with a bin-dependent factor rather than summing up the energy of the individual hits. In this way, Landau fluctuations are suppressed, motivated by the assumption that these hits primarily originate from a single or few particles.
Then, each $E_i$ is multiplied with a bin-dependent weight and the total reconstruction energy is determined by the  sum of the bin energies together with the contribution of the primary track (before the shower starts), which is reconstructed with the standard reconstruction.

\begin{figure}[h]
\includegraphics[width = 0.5\textwidth]{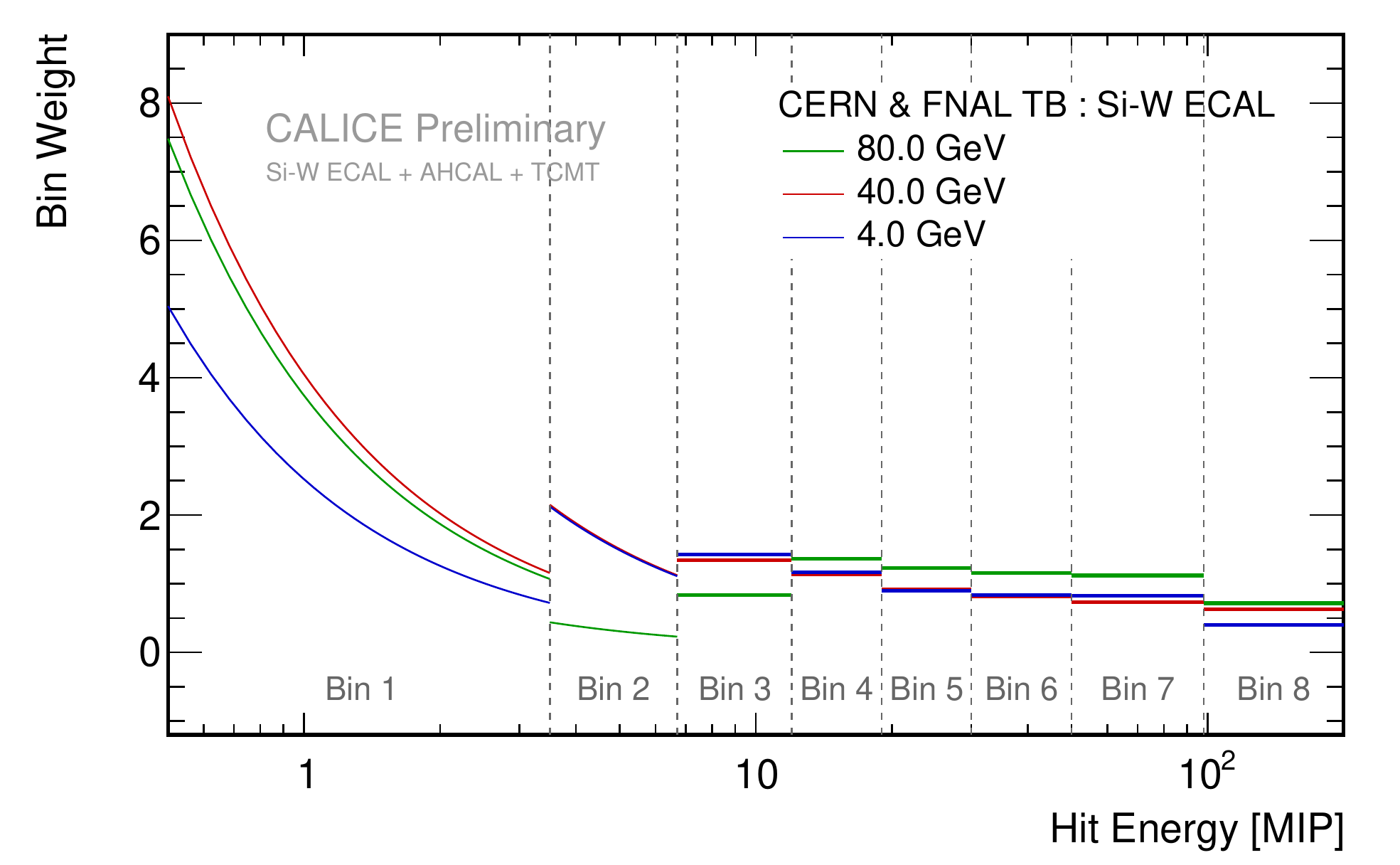} 
\hfill
\includegraphics[width = 0.5\textwidth]{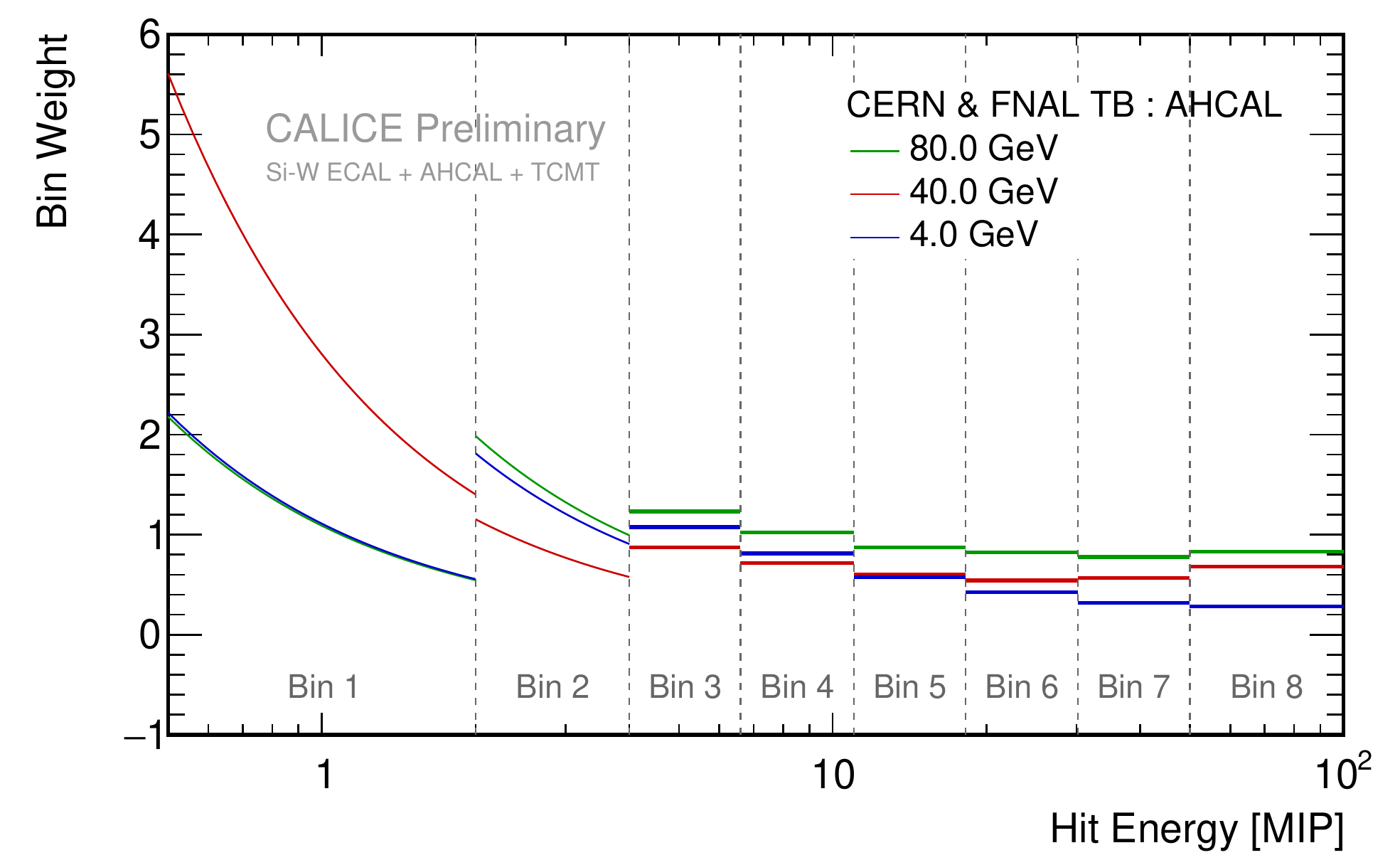} 
\caption{The optimized bin weights of the Si-W ECAL (left) and AHCAL (right) for 4\,GeV, 40\,GeV and 80\,GeV $\pi^-$ beams. The use of  a constant value rather than the individual amplitude of each hit in the first two bins yields   energy dependent weights in these bins.  }
\label{fig:Weights}
\end{figure}

The bin-dependent weights, which are applied to the energy sums in the different bins, are  energy dependent due to the energy dependence of the shower density profile as well as  the average electromagnetic fraction of the shower. A parametrization of  second order polynomials as a function of the particle energy is used to describe this dependence, resulting in   three   parameters for each bin. In total, this results in  51 parameters which are needed to calculate the software compensation weights for the full system (24 each for the ECAL and the AHCAL, 3 for the TCMT). These parameters are being optimized by the minimization of  a $\chi^2$ function calculated over  multiple runs at multiple energies using the appropriated beam energies. 
After the optimization,  these 51 parameters are implemented in the corresponding polynomials, using the  corresponding standard reconstructed energy as an input in order to reconstruct each test beam energy. In that way, no prior knowledge of the beam energy is used when applying the SC reconstruction to the data.      
In figure \ref{fig:Weights}, the optimized weights for the Si/Scint setup for 4\,GeV, 40\,GeV and 80\,GeV  $\pi^-$ beams are presented.
The higher energy bins, which are assumed to contain mostly hits of electromagnetic sub-showers, tend to be weighted below unity while the  lower energy bins  tend for weights  above unity. 

\begin{figure}[h]
\begin{center}
\includegraphics[width = 0.7\textwidth]{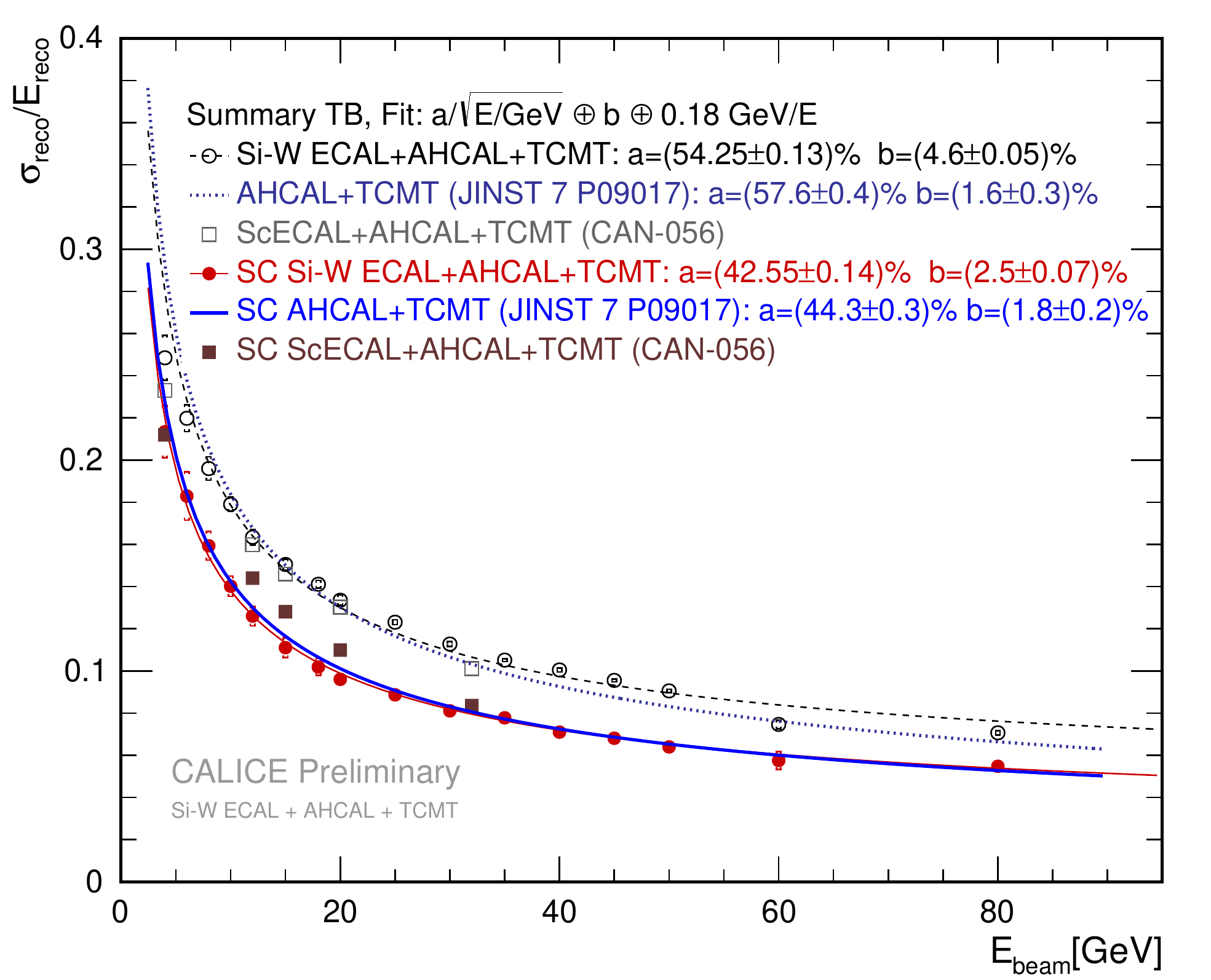} 
\end{center}
\caption{The reconstructed energy resolutions with the standard and Full SC methods for the Si/Scint and the all-Scint setups  in comparison with the energy resolution fits published in \protect\cite{Frank}.  The fit parameters are given in the legend, where the noise term is fixed to $c$ = 0.18\,GeV, as was determined from  noise measurements in dedicated runs without beam particles as well as in random trigger events  in the CERN test beam setup \cite{Frank}. 
}
\label{fig:Results}
\end{figure}

The energy resolutions obtained with the standard  and the SC reconstruction methods   are presented in figure \ref{fig:Results}  for the Si/Scint and the all-Scint setups. The SC method achieves an overall improvement in the energy resolution in both of the setups, with  a maximum relative improvement of 20\% in the all-Scint setup and 30\% in the Si/Scint setup. The latter can be seen also from the decrease of the stochastic term of the fits from (54.25$\pm$0.13)\% to (42.55$\pm$0.14)\% with the SC reconstruction.

The figure also includes  the resolution fits shown in \cite{Frank} with the standard reconstruction and the\textit{ local SC} method, where only events with showers starting in the AHCAL are included and the SC method is applied merely to the AHCAL and the TCMT. The standard reconstruction of the full systems is compatible with the AHCAL contained system till 20\,GeV, above which  a slightly worse resolution is observed in the combined systems. However, with the SC reconstruction, the full calorimeter systems perform similarly to the previous results, despite of the difference in the geometry and the sampling structure of the sub-detectors, and the different readout technology in the Si/Scint setup.

\section{SC Application  to different Detectors}

\begin{figure}[h]
\includegraphics[width = 0.5\textwidth]{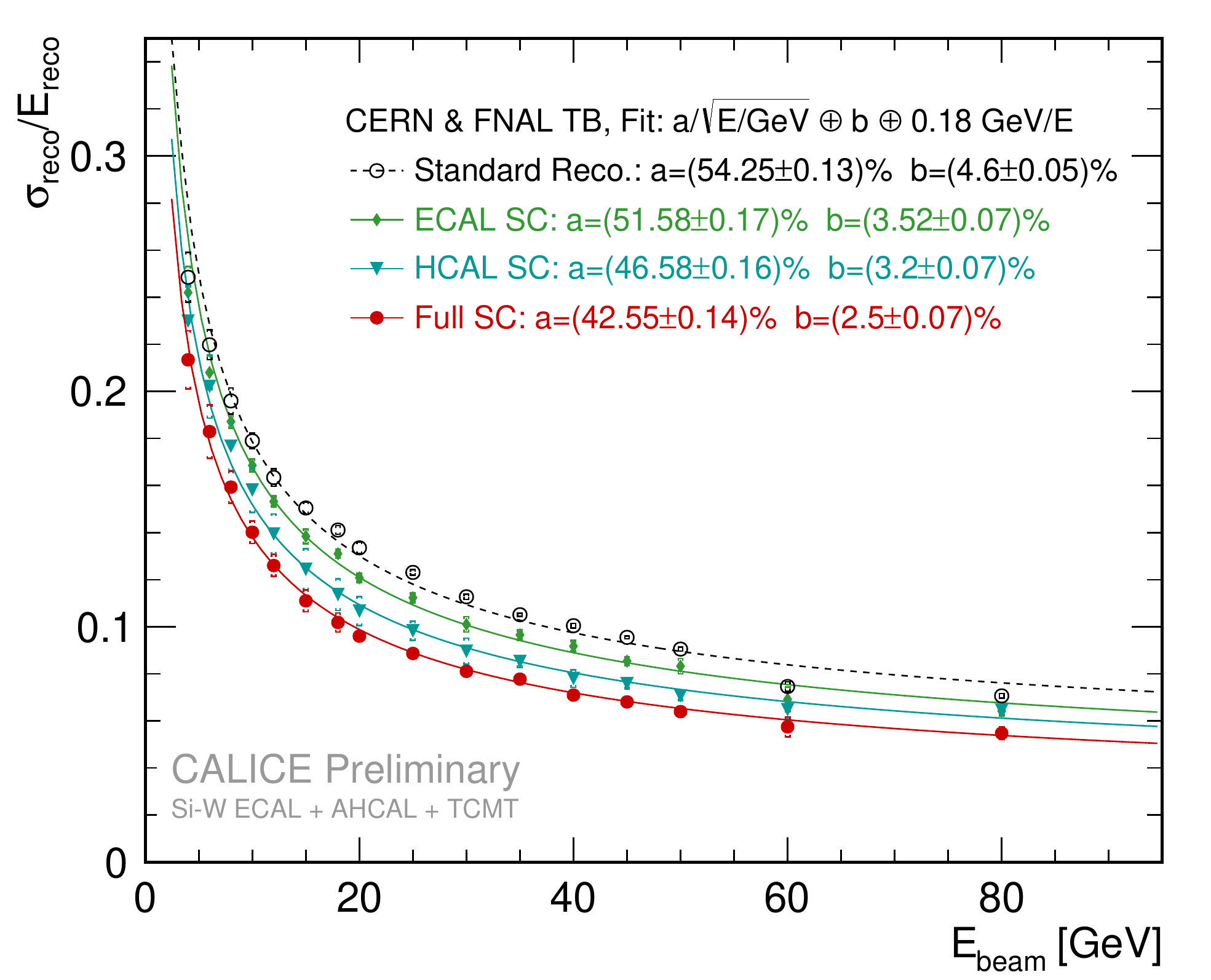} 
\hfill
\includegraphics[width = 0.5\textwidth]{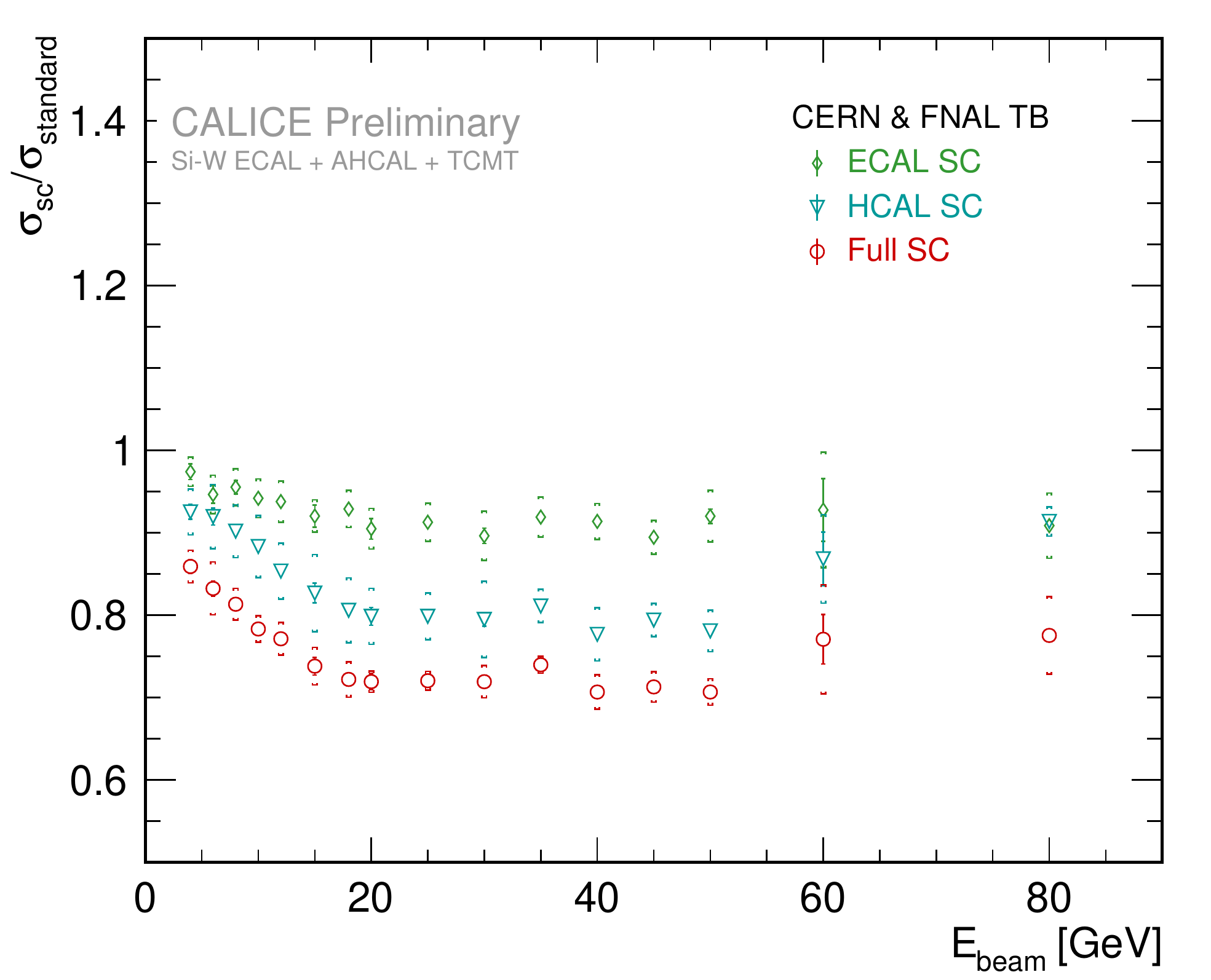} 
\caption{Reconstructed energy resolutions( left) and relative improvement of the resolutions (right) with standard, ECAL SC, AHCAL SC and Full SC reconstruction methods for the Si/Scint setup.  The fit parameters are given in  the legend. The total (statistical and systematic) uncertainties are marked with '[]'.}
\label{fig:resDetectors}
\end{figure}

The SC scheme discussed so far includes all the three sub-detectors in the combined setups and therefore is referred to as the \textit{Full SC} scheme. 
To study the application of SC reconstruction to individual parts of the combined systems, two additional SC schemes were used to reconstruct the data of  Si/Scint setup: the \textit{ECAL SC}, where SC is applied only to the Si-W ECAL and the \textit{HCAL SC}, where SC is applied to the AHCAL and  the TCMT.
These schemes use the same binning technique for the corresponding sub-detectors as the Full SC scheme, while the contributions of the primary track and the additional sub-detectors are reconstructed with the standard method. 
Figure  \ref{fig:resDetectors} presents the energy resolutions  with the standard  and the SC reconstruction methods  and the relative improvement of the resolutions with the SC reconstruction methods compared to the standard reconstruction ($\sigma_\text{reco}/\sigma_\text{sc}$). 

The  improvement of the resolution with the SC techniques is energy dependent  with maximums of only a few percent with ECAL SC, approximately 22\% with  AHCAL SC  and approximately 30\% with the Full SC.
%
This effect can be seen as well from the reduction in the stochastic term of the energy resolution fits from (54.25$\pm$0.13)\% with the standard reconstruction to  (51.58$\pm$0.17)\%, (46.58$\pm$0.16)\% and (42.55$\pm$0.14)\%   with the ECAL SC, the HCAL SC and the Full SC, respectively.

\section{Application of Weights to Different Beam Periods}

A systematic test for the robustness of the SC method was performed with the data recorded with the Si/Scint setup in CERN and FNAL test beam experiments. The data of each experiment was reconstructed separately with the Full SC method, where a set of SC weights for the respective test beam was trained.
Then, the SC weights obtained from CERN data were implemented in the fitting energy range on FNAL data  and vice versa.
This  gives a realistic evaluation of the changes in the setup performance between the two experiments due to  different beam configurations, hardware changes  (repairs and changes of detector elements and electronics) and after reassembly of the calorimetry system.

\begin{figure}[h]
\includegraphics[width = 0.495\textwidth]{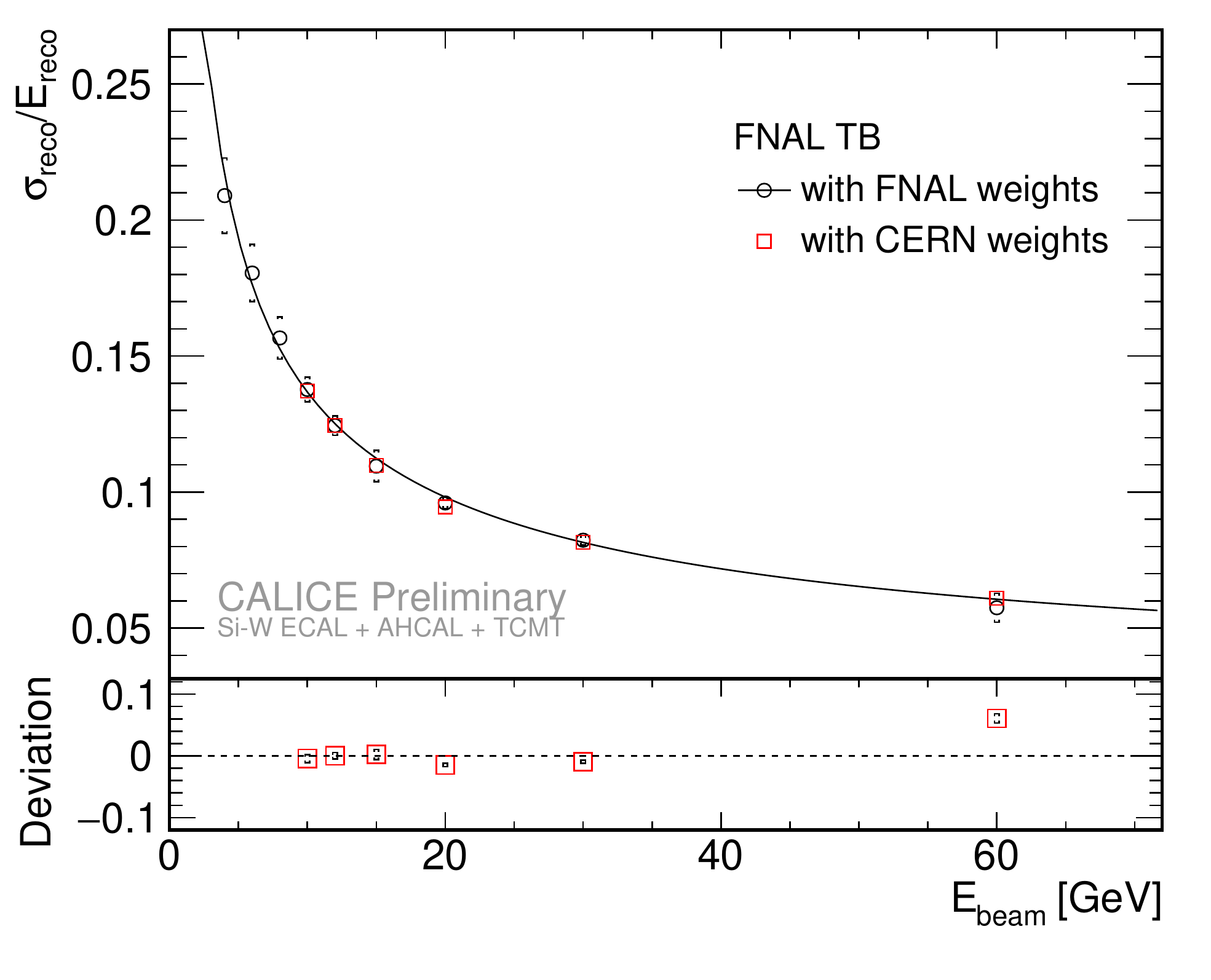} 
\hfill
\includegraphics[width = 0.495\textwidth]{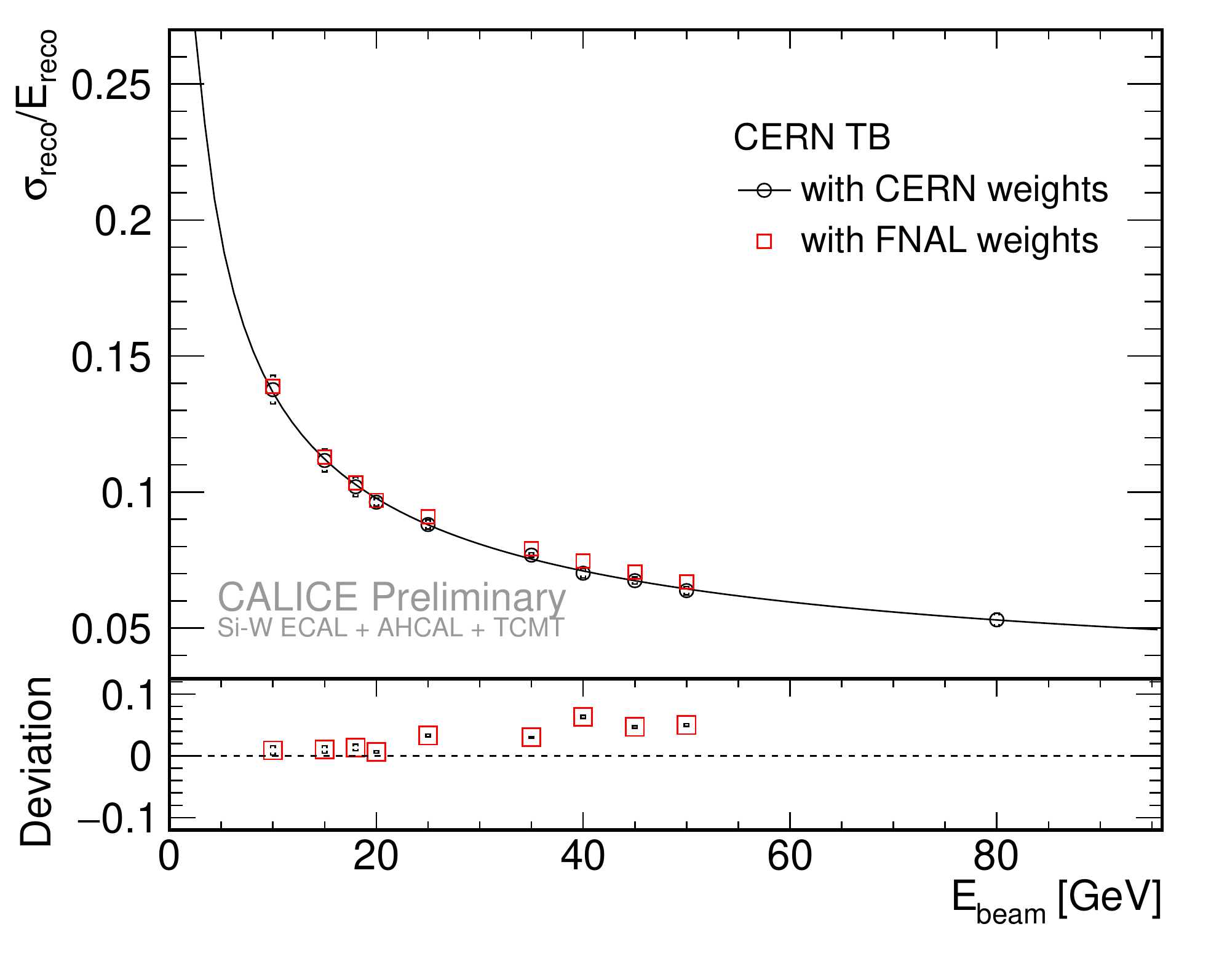} 
\caption{
Reconstructed energy resolutions  using  SC weights optimized for the respective dataset (black) and  SC weights trained for a different dataset (red) and the relative deviations between them.
The total (statistical and systematic) uncertainties are marked with '[]'. }
\label{fig:CERN_wFNAL}
\end{figure}

 In figure \ref{fig:CERN_wFNAL} the difference in the reconstructed energy resolutions is shown.  For FNAL data with CERN weights, the deviations in the performance are negligible excluding the 60\,GeV data point, where a deterioration of 6\% is observed. For CERN data with FNAL weights, there is an energy dependent deterioration, rising up to 6\% at 40\,GeV. In both cases, using different data for training the weights produces a relatively small effect on the energy resolutions and thus demonstrates the robustness of the SC method.

\section{Conclusions}
The CALICE collaboration has installed and tested combined ECAL and HCAL systems in test beam experiments to study  the system performance in  realistic detector configurations. 
The combined Si-W ECAL, AHCAL and TCMT and the combined ScECAL, AHCAL and TCMT have recorded 
data of charged pions in the total energy range of 4\,GeV to 80\,GeV. 
These data were reconstructed with the standard and the SC method and a similar performance as shown in a previous study focused on  AHCAL data was observed. This demonstrates the success of the developed calibration process for combined systems which include different geometries and readout technologies. 

The SC reconstruction method was applied to the full system as well as to individual calorimeters and shows an overall improvement  of the energy resolution compared to the standard reconstruction method. This improvement is maximal when the SC scheme was applied to the full system, rising up to approximately 30\%  with the Si/Scint setup and approximately 20\%  with the all-Scint setup.



%
%
%


\bibliography{Energy_Reconstruction_of_Hadrons_in_highly_granular_combined_ECAL_and_HCAL_systems}
\bibliographystyle{unsrt}
\clearpage
%
%



\end{document}